\documentclass[letterpaper]{article} 
\usepackage{aaai2026}  
\usepackage{times}  
\usepackage{helvet}  
\usepackage{courier}  
\usepackage[hyphens]{url}  
\usepackage{graphicx} 
\usepackage{natbib}  
\usepackage{caption} 
\usepackage{algorithm}

\usepackage{newfloat}
\usepackage{listings}
\DeclareCaptionStyle{ruled}{labelfont=normalfont,labelsep=colon,strut=off} 
\floatstyle{ruled}
\newfloat{listing}{tb}{lst}{}
\floatname{listing}{Listing}
\usepackage[utf8]{inputenc} 
\usepackage{url}            
\usepackage{booktabs}       
\usepackage{amsfonts}       
\usepackage{nicefrac}       
\usepackage{microtype}      
\usepackage{xcolor}         
\usepackage{xspace}
\usepackage{threeparttable}
\usepackage{adjustbox}
\usepackage{amsmath}
\usepackage{multirow}
\usepackage{enumitem}
\usepackage{algpseudocode}

\def\fig{Fig.\xspace}
\def\eqn{Eq.\xspace}

\def\tab{Tab.\xspace}
\def\alg{Alg.\xspace}
\def\apx{Apx.\xspace}

\def\sysname{Apo2Mol\xspace}

\ifodd 1
\newcommand{\com}[1]{\textbf{\color{red}(COMMENT: #1)}} 
\newcommand{\todo}[1]{\textbf{{\color{orange}(TODO: #1)}}}
\else

\newcommand{\com}[1]{}
\newcommand{\todo}[1]{}
\fi

\title{\sysname: 3D Molecule Generation via Dynamic Pocket-Aware Diffusion Models}
\author{
    Xinzhe Zheng\textsuperscript{\rm 1}, 
    Shiyu Jiang\textsuperscript{\rm 1},
    Gustavo Seabra\textsuperscript{\rm 1},
    Chenglong Li\textsuperscript{\rm 1},
    Yanjun Li\textsuperscript{\rm 1,2,\thanks{Corresponding author: yanjun.li@ufl.edu}}
}
\affiliations{
    \textsuperscript{\rm 1}Department of Medicinal Chemistry, Center for Natural Products, Drug Discovery and Development, \\ University of Florida, Gainesville, FL, USA\\
    \textsuperscript{\rm 2}Department of Computer and Information Science and Engineering, University of Florida, Gainesville, FL, USA
}

\usepackage{bibentry}

\begin{document}

\maketitle

\begin{abstract}
Deep generative models are rapidly advancing structure-based drug design, offering substantial promise for generating small molecule ligands that bind to specific protein targets.
However, most current approaches assume a rigid protein binding pocket, neglecting the intrinsic flexibility of proteins and the conformational rearrangements induced by ligand binding, limiting their applicability in practical drug discovery.
Here, we propose \sysname, a diffusion-based generative framework for 3D molecule design that explicitly accounts for conformational flexibility in protein binding pockets.
To support this, we curate a dataset of over $24,000$ experimentally resolved apo-holo structure pairs from the Protein Data Bank, enabling the characterization of protein structure changes associated with ligand binding.
\sysname employs a full-atom hierarchical graph-based diffusion model that simultaneously generates 3D ligand molecules and their corresponding holo pocket conformations from input apo states.
Empirical studies demonstrate that \sysname can achieve state-of-the-art performance in generating high-affinity ligands and accurately capture realistic protein pocket conformational changes.
\end{abstract}

\begin{links}
    \link{Code}{https://github.com/AIDD-LiLab/Apo2Mol}
\end{links}

\section{Introduction}
\label{sec:intro}

Structure-based drug design (SBDD) leverages the three-dimensional structures of protein targets to guide the rational therapeutic design~\cite{anderson2003process}.
By capturing the geometry and physicochemical features of binding sites, SBDD enables the generation of ligands with high affinity and specificity.
The advances of deep learning have further catalyzed the SBDD field, with deep generative models showing remarkable success in generating ligands tailored to specific pockets~\cite{zhang2025structure}.
Early studies primarily operated on 1D string-based or 2D graph-based molecular representations, while subsequent approaches advanced to directly generating 3D ligand conformations~\cite{bilodeau2022generative}.
More recently, diffusion models have emerged as a powerful framework, iteratively denoising atoms within the pocket to yield novel ligands together with their 3D poses~\cite{guan2024decompdiff,schneuing2024structure,li2025molecule,lin2025diffbp}.

\begin{figure}[!t]
    \centering
    \includegraphics[width=0.46\textwidth]{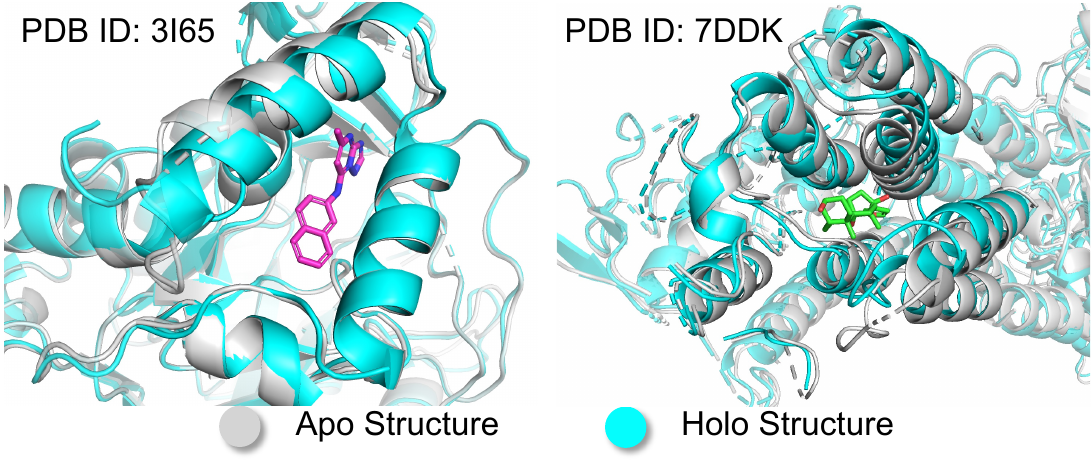}
    \caption{
        Conformational changes between the apo (unbound) and holo (ligand-bound) states, illustrating ligand-induced structural dynamics.
    }
    \label{fig:intro-example}
\end{figure}

Despite these advances, a fundamental limitation persists in most current approaches to SBDD: the assumption of a static, rigid protein binding pocket~\cite{guan2024decompdiff,schneuing2024structure,li2025molecule}.
Modeling the receptor pocket as rigid neglects the dynamic nature of molecular recognition (\fig~\ref{fig:intro-example}), biasing design toward a single holo (ligand-bound) conformation and potentially missing ligands that would induce alternative, more favorable pocket geometries~\cite{wei2024structure}.
This limitation becomes especially acute for emerging targets or those for which only apo (unbound) conformations are available, as current SBDD methods often struggle to generate high-quality ligands in the absence of an appropriate bound-state template.
A recent work DynamicFlow~\cite{zhou2025integrating} relaxes the rigidity assumption by introducing a full-atom stochastic flow-matching model trained on molecular dynamics (MD) simulation trajectories from the MISATO dataset~\cite{siebenmorgen2024misato}, enabling joint generation of the holo pocket conformation and the bound ligand from an initial apo structure.
While promising, MD simulations are computationally expensive and sensitive to force field parameters and timescales~\cite{hollingsworth2018molecular}. 
Moreover, simulated dynamics may fail to capture experimental conformational transitions, potentially introducing simulation-specific artifacts into generative models~\cite{hollingsworth2018molecular}.

To overcome the limitations of rigid-pocket assumptions and the dependency on potentially biased simulation data, we introduce \sysname, a diffusion-based full-atom framework for SBDD that explicitly considers the dynamics of the protein binding pocket (see \fig\ref{fig:sys-overview}).
We address two central challenges in tandem: (1) grounding training on a large-scale dataset derived exclusively from experimentally resolved structures, and (2) modeling 3D ligand generation with protein structural dynamics.
Instead of relying on MD simulations, we curate a high-quality dataset of experimentally resolved apo-holo protein structure pairs stored in the Protein Data Bank (PDB) by leveraging the PLINDER resource~\cite{durairaj2024plinder}.
Rigorous filtering criteria (\apx A) is applied to retain pharmacologically relevant ligands, high-resolution structures ($\leq 2.5$Å), and apo-holo pairs with $100\%$ sequence identity.
This yields a final dataset comprising $24,601$ apo-holo pairs.
Compared with ApoBind~\cite{aggarwal2021apobind}, which contains $\sim10K$ samples with $\ge80\%$ apo-holo sequence identity, our dataset is substantially larger and better aligned with the requirements of this study.
Building upon our curated dataset, \sysname employs a diffusion-based framework in which the forward diffusion gradually perturbs the ligand coordinates and atom types, and simultaneously interpolates protein pocket coordinates from holo towards apo conformations.
During reverse diffusion, the apo pocket serves as the initial structural condition, the model learns to denoise a randomly initialized ligand while concurrently transforming the pocket conformation from apo to its holo state.
To better capture ligand–pocket interactions, \sysname integrates SE(3)-equivariant attention layers within a hierarchical graph-based message-passing framework that aggregates atom-level information into residue-level representations, enabling more expressive and spatially aware modeling.
Our experimental results demonstrate that \sysname generates high-affinity ligands and produces reasonable pocket conformational changes, advancing the capabilities of dynamic SBDD beyond static structure assumptions.

\section{Related Work}
\label{sec:related-work}

\subsection{Structure-based Drug Design}
Structure-based drug design leverages the three-dimensional structures of protein targets to rationally design small molecules with desired binding affinities. Early approaches generated 1D strings, e.g. SMILES~\cite{weininger1988smiles}, or 2D molecular graphs conditioned on protein contexts~\cite{skalic2019target,atz2024prospective,tan2023target,wang2025improving}.
Subsequent methods employed variational auto encoders (VAEs) and autoregressive models to directly produce 3D ligand conformations aligned with receptor binding sites~\cite{liu2022generating,Pocket2Mol,ragoza2022generating,zhang2023molecule}.
Recent diffusion-based approaches further advanced SBDD by iteratively denoising atom types and positions within the protein pocket, followed by post-processing bond formation~\cite{guan3d, lin2025diffbp,schneuing2024structure}.
DecompDiff~\cite{guan2024decompdiff} introduces the decomposition of biochemical priors into scaffold and arm structures.
IPDiff~\cite{huang2024protein} further improves upon this by explicitly incorporating binding-aware biochemical priors. Hierarchical consistency diffusion is adopted to integrate atom- and motif-level structural priors for consistent generation~\cite{li2025molecule}.
DynamicFlow~\cite{zhou2025integrating} proposes integrating protein dynamics into SBDD through training on MD trajectories, using a full-atom stochastic flow model.
While existing \textit{de novo} design methods are trained exclusively on static holo structures or simulated dynamics, our \sysname explicitly leverages experimentally validated apo-holo structure pairs to jointly denoise protein pockets and ligands, effectively capturing ligand-binding dynamics.

\subsection{Modeling Protein-ligand Interaction Dynamics}
Proteins are inherently dynamic, and their functional states often involve conformational changes upon ligand binding.
Understanding the dynamic process is crucial for applications like drug discovery and design~\cite{wei2024structure,li2022dyscore}.
Molecular dynamics simulations have traditionally served as the cornerstone for exploring protein conformational landscapes and transition pathways at atomic detail~\cite{alonso2006combining, dror2012biomolecular, hollingsworth2018molecular}.
Recently, machine learning--especially deep generative models--has shown great promise in modeling dynamics and generating conformational ensembles for protein and small molecule~\cite{janson2023direct, kuznetsov2024cosmic, lu2024structure,wang2024protein}, as well as protein-ligand docking~\cite{corsocomposing, lu2024dynamicbind, morehead2024flowdock, morehead2025deep}.
While the above modeling approaches have incorporated conformational dynamics, such considerations are largely absent from most \textit{de novo} small molecule generation methods~\cite{grisoni2023chemical, tang2024survey,xie2022advances}.
In this work, \sysname jointly generates both the ligand molecules and the corresponding holo pocket conformations from apo states, capturing the inherent dynamics of molecular interaction within a unified generative framework.

\begin{figure*}[!t]
    \centering
    \includegraphics[width=0.85\textwidth]{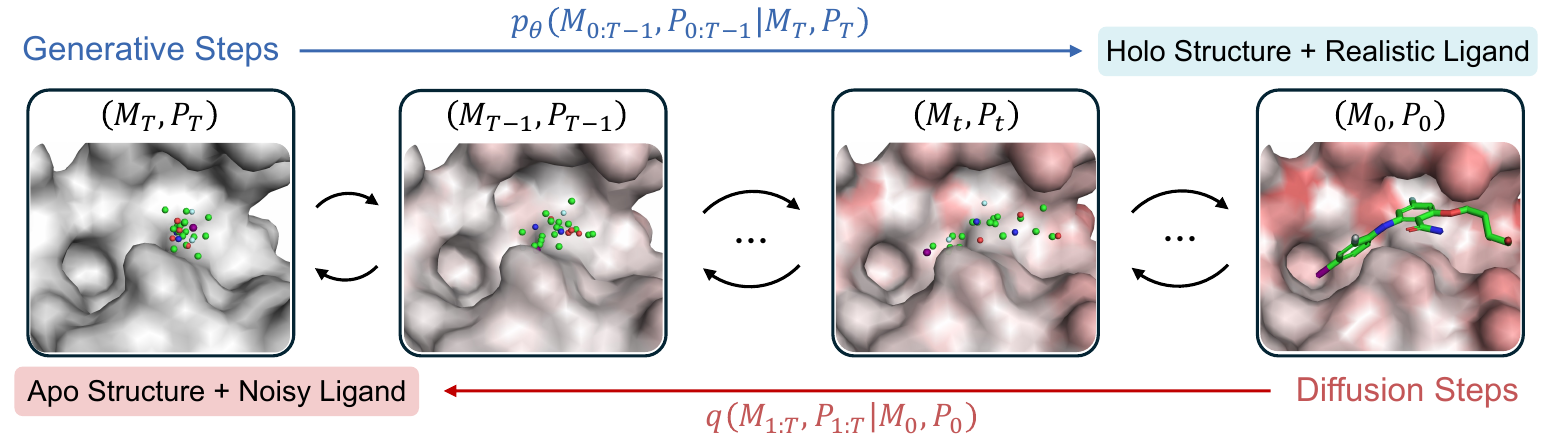}
    \caption{Schematic overview of \sysname. The diffusion process gradually corrupts an experimental holo pocket–ligand pair by injecting noise into the ligand and linearly interpolating the pocket from its holo conformation toward the apo state. The generative process learns to denoise the corrupted inputs and recover the joint distribution of the holo pocket conformations and their corresponding ligand types and poses. Structural deviations between the apo and holo conformations are illustrated in red.}
    \label{fig:sys-overview}
\end{figure*}

\section{Preliminary}
\label{sec:preliminary}

\subsection{Problem Statement}
In this work, we address the challenge of jointly generating a holo protein pocket and its corresponding rational ligand from an apo protein conformation.
The holo protein pocket is represented as $\mathcal{P}_{\text{H}}=\{\boldsymbol{x}^{i}_{\mathcal{P}_{\text{H}}}, \boldsymbol{v}^{i}_{\mathcal{P}_{\text{H}}}\}_{i=1}^{N_{\mathcal{P}}}$, and its corresponding apo conformation and the generated small molecule are denoted as
$\mathcal{P}_{\text{A}}=\{\boldsymbol{x}^{i}_{\mathcal{P}_{\text{A}}}, \boldsymbol{v}^{i}_{\mathcal{P}_{\text{A}}}\}_{i=1}^{N_{\mathcal{P}}}$ and $\mathcal{M}=\{\boldsymbol{x}^{i}_{\mathcal{M}}, \boldsymbol{v}^{i}_{\mathcal{M}}\}$, respectively.
Here, $N_{\mathcal{P}}$ and $N_{\mathcal{M}}$ represent the number of atoms in the protein pocket and the ligand.
$\boldsymbol{x}^{i} \in \mathbb{R}^3$ refers to the 3D atom coordinates, while $\boldsymbol{v}^{i} \in \mathbb{R}^K$ represents the atom features.
The learning objective can be formulated as modeling the conditional distribution of $p(\mathcal{P}_{\rm{H}},\mathcal{M} | \mathcal{P}_{\rm{A}})$.

For the ligand generation, the model directly predicts the ligand atom types and associated 3D coordinates.
In contrast, for holo pocket prediction, we refrain from estimating atomic coordinates to preserve the structural integrity of the protein pocket.
Following prior works~\cite{lu2024dynamicbind, zhang2023diffpack}, we model pocket conformational changes at the residue level by predicting translations $\boldsymbol{tr} \in \mathbb{R}^3$, rotations $\boldsymbol{q} \in \mathbb{H}$, and chi angle updates $\boldsymbol{\mathcal{X}} = \{\mathcal{X}_i \in \mathrm{SO}(2)\}_{i=0}^4$, which correspond to side-chain torsion angles.
To represent 3D rotations, we adopt quaternions $\boldsymbol{q}$, which offer superior numerical stability over rotation vectors, eliminate singularities such as gimbal lock, and enable smooth interpolation~\cite{diebel2006representing}.
This formulation enables physically plausible transformations between apo and holo states.

\subsection{Diffusion Models}
Diffusion models~\cite{ho2020denoising, song2020score} are a class of generative models that have achieved remarkable success in multiple domains~\cite{zhu2023conditional, epstein2023diffusion,vignac2022digress, hoogeboom2022equivariant}.
These models are inspired by stochastic differential equations (SDEs) that transform a complex data distribution $p_{\text{data}}$ into a simple prior distribution, typically the Gaussian distribution, through the gradual noise injection:
\begin{equation}
    \mathrm{d} \mathbf{x}_t = \boldsymbol{\mu}(\mathbf{x}_t, t)\,\mathrm{d}t + \sigma(t)\,\mathrm{d}\mathbf{w}_t,
\end{equation}
where $t\in[0, T]$, $T>0$ is a fixed constant, $\mathbf{w}_t$ denotes Brownian motion, and $\boldsymbol{\mu}$, $\sigma$ are drift and diffusion coefficients, respectively.
A notable property of this process is the existence of a deterministic reverse-time ordinary differential equation (ODE), called the probability flow ODE, whose solution follows the same marginal $p_t$:
\begin{equation}
    \frac{\mathrm{d} \mathbf{x}_t}{\mathrm{d}t} = \boldsymbol{\mu}(\mathbf{x}_t, t) - \frac{1}{2}\sigma(t)^2 \nabla_{\mathbf{x}} \log p_t(\mathbf{x}_t),
    \label{eqn:pfode}
\end{equation}
where $\nabla \log p_t(\mathbf{x})$ is the score function. When $\boldsymbol{\mu} = 0$ and $\sigma(t) = \sqrt{2t}$, this becomes a pure Gaussian diffusion, and $p_t$ remains analytically tractable.

To estimate the score term, diffusion models employ a neural network $\boldsymbol{s}_\theta(\mathbf{x}, t) \approx \nabla \log p_t(\mathbf{x})$, trained using score matching to obtain an empirical estimate of the probability flow ODE:
\begin{equation}
    \frac{\mathrm{d} \mathbf{x}_t}{\mathrm{d}t} = -t\boldsymbol{s}_\theta(\mathbf{x}, t).
\end{equation}
Once trained, samples can be generated by solving \eqn\ref{eqn:pfode} backward in time, typically starting from a Gaussian prior and integrating to $t = \epsilon > 0$ for numerical stability.
This yields an approximate sample from the data distribution $p_{\text{data}}$.

\section{Method}
\label{sec:method}

In this work, we propose \sysname, a framework that captures the dynamic nature of protein pockets for SBDD.
As illustrated in \fig~\ref{fig:sys-overview}, \sysname interpolates the protein pocket between its apo and holo conformations, generating interaction pairs by combining these intermediate pocket states with noisy ligands at each time step.
The red regions on the protein pocket highlight conformational deviations from the apo structure.
The following section presents a detailed overview of the \sysname framework.

\subsection{Data Preparation}

\paragraph{Pocket Alignment}
We first apply root-mean-square deviation (RMSD) alignment to superimpose the apo and holo pocket structures in 3D space.
Then, the Kabsch algorithm~\cite{kabsch1976solution} is utilized for calculating the residue-level translations $\boldsymbol{tr}$ and rotations $\boldsymbol{q}$, centered on the $C_{\alpha}$ atoms, that align the holo backbone atoms $N\text{-}C_{\alpha}\text{-}C$ to the apo structure:
$\boldsymbol{tr}, \boldsymbol{q} = \text{Kabsch}(\boldsymbol{x}_{N\text{-}C_{\alpha}\text{-}C}^{\text{holo}} - \boldsymbol{x}_{N\text{-}C_{\alpha}\text{-}C}^{\text{apo}})$.
Finally, using the utilities provided in DynamicBind~\cite{lu2024dynamicbind}, we extract the chi angles for both holo and apo structures, and calculate the chi angle update as:$\boldsymbol{\mathcal{X}}=\boldsymbol{\mathcal{X}}^{\text{apo}}-\boldsymbol{\mathcal{X}}^{\text{holo}}$.

\begin{figure*}[!t]
    \centering
    \includegraphics[width=0.85\textwidth]{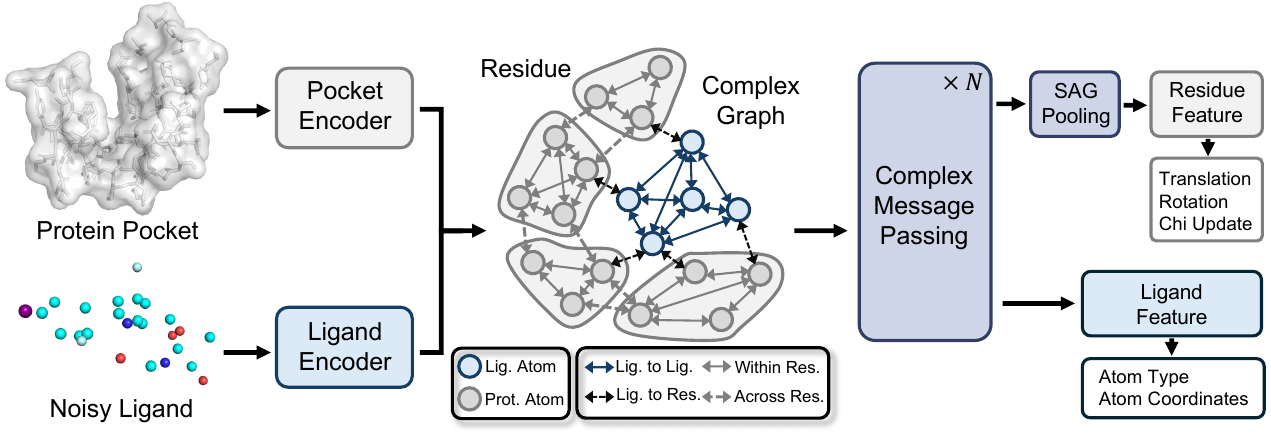}
    \caption{Illustration of the \sysname framework, which jointly models ligand generation and protein pocket refinement through hierarchical graph message passing.}
    \label{fig:network-pipeline}
\end{figure*}

\paragraph{Pseudo-interaction Pair}
Given a paired protein pocket and its corresponding ligand, we construct the pseudo-interaction pair as the model input at each diffusion step $t$.

For the ligand, we adopt the noise injection scheme from prior works~\cite{huang2024protein, huang2024interaction}, where both position and atom-type representations are corrupted progressively.
At each time step $t$, the forward process is defined as:
\begin{equation}
\label{eqn:ligand-t}
\begin{aligned}
    q(\boldsymbol{x}_t^{\mathcal{M}} \mid \boldsymbol{x}_0^{\mathcal{M}})&=\mathcal{N}(\boldsymbol{x}_t^{\mathcal{M}};\sqrt{\overline{\alpha}_t}\boldsymbol{x}_0^{\mathcal{M}}, (1-\overline{\alpha}_t)\mathbf{I}), \\
    q(\boldsymbol{v}_t^{\mathcal{M}} \mid \boldsymbol{v}_0^{\mathcal{M}}) &= \mathcal{C}\left(\boldsymbol{v}_t^{\mathcal{M}} \mid \overline{\alpha}_t\boldsymbol{v}_{0}^{\mathcal{M}} + (1-\overline{\alpha}_t)/K  \right),
\end{aligned}
\end{equation}
where $\mathcal{N}$ and $\mathcal{C}$ denote the Gaussian and categorical distributions respectively.
The scalar $\alpha_t$ is determined by a predefined noise schedule.

Unlike the ligand, the conformation of the protein pocket at step $T$ is not sampled from a fixed prior but rather corresponds to the apo state.
To simulate intermediate pocket conformations along the apo–holo trajectory, we adopt a residue-level interpolation strategy that ensures structural coherence.
Specifically, we interpolate residue-level translations, rotations, and chi angles updates.
For translations and chi angles, the interpolated states at time $t$ are modeled as:
\begin{equation}
\begin{aligned}
\label{eqn:tr-x-t}
\boldsymbol{tr}_{t} \sim \mathcal{N}((t/T)\boldsymbol{tr}_0, \lambda_{tr} \beta_t \mathbf{I}), \\
\boldsymbol{\mathcal{X}}_{t} \sim \mathcal{N}\left((t/T)\boldsymbol{\mathcal{X}}_0, \lambda_{\mathcal{X}} \beta_t \mathbf{I} \right),
\end{aligned}
\end{equation}
where $\lambda_{tr}$ and $\lambda_{\mathcal{X}}$ control the noise scale applied to translations and chi angles, respectively.
$\beta_t$ is equal to $1-\alpha_t$.

For rotations, which are represented as quaternions, we use spherical linear interpolation (Slerp)~\cite{shoemake1985animating} to smoothly transition from the holo-to-apo rotation $\boldsymbol{q}$ to the unit quaternion $\boldsymbol{q}_{\text{unit}}$ (no rotation):
\begin{equation}
\label{eqn:q-t}
\boldsymbol{q}_{t} = \text{Slerp}\left(\boldsymbol{q}_{\text{unit}}, \boldsymbol{q}_0, t/T \right) \otimes \boldsymbol{\epsilon}_t,
\boldsymbol{\epsilon}_t \sim \mathcal{N}_{\mathbb{H}}\left(\mathbf{1}, \lambda_{q} \beta_t \right),
\end{equation}
where $\otimes$ represents quaternion multiplication, and $\mathcal{N}_{\mathbb{H}}(\mathbf{1}, \lambda_q \beta_t)$ is a perturbation distribution centered at the identity quaternion with variance scaled by $\lambda_q \beta_t$.
Noise is added to improve robustness, helping the model escape local minima and generalize across diverse conformational pathways.

The pocket coordinates at time $t$ are then computed by applying the interpolated transformations to the holo structure:
\begin{equation}
    \boldsymbol{x}_t^{\mathcal{P}} = (\boldsymbol{\mathcal{X}}_t)(\boldsymbol{q}_t \boldsymbol{x}_0^{\mathcal{P}}\boldsymbol{q}_t^{*} + \boldsymbol{tr}_t),
\end{equation}
where $\boldsymbol{q}_t^{*}$ is the conjugate version of the quaternion.

Finally, by assembling the components described above, we obtain the pseudo interaction pair as $(\boldsymbol{x}_t^{\mathcal{M}},\boldsymbol{v}_t^{\mathcal{M}};\boldsymbol{x}_t^{\mathcal{P}}, \boldsymbol{v}_t^{\mathcal{P}})$.

\subsection{\sysname Framework}
\fig~\ref{fig:network-pipeline} illustrates the network architecture of \sysname.
The following content provides details about each component.

\paragraph{Feature Extraction}
The protein pocket and the noisy ligand at time step $t$ are encoded separately.
Ligand atoms are one-hot encoded by atom type, while protein atoms are represented by three one-hot vectors that specify atom type, amino acid identity, and a backbone indicator.
These features are projected into a shared $d$-dimensional embedding space using two-layer multilayer perceptrons (MLPs).
We denote the resulting embeddings as $\boldsymbol{h}^{\mathcal{M}} \in \mathbb{R}^{N_{\mathcal{M}} \times d}$ for the ligand and $\boldsymbol{h}^{\mathcal{P}} \in \mathbb{R}^{N_{\mathcal{P}} \times d}$ for the pocket.

\paragraph{Complex Graph Construction}
After obtaining the hidden features, \sysname constructs a protein-ligand complex graph to enable graph neural network (GNN) training.
A $k$-nearest neighbor strategy is employed to establish edges based on spatial proximity, connecting nodes from both the protein pocket and the ligand.
To facilitate learning of meaningful protein-ligand interaction, while capturing residue-level conformational dynamics of the pocket, four distinct types of edges are defined: 1) intra-ligand edges; 2) ligand–residue edges; 3) intra-residue edges, and 4) inter-residue edges across different residues.
This hierarchical design provides a multi-scale representation that supports accurate modeling of ligand binding and pocket refinement.

\paragraph{Complex Message Passing}
To capture intricate molecular interactions during the generative process, \sysname employs a SE(3)-equivariant~\cite{satorras2021n} GNN model with attention mechanism over the constructed protein–ligand complex graph.
Each node in the graph is associated with both: a 3D equivariant feature $\boldsymbol{x} \in \mathbb{R}^3$, representing the spatial position, and a 3D invariant feature $\boldsymbol{h} \in \mathbb{R}^d$ of an atom, representing its latent chemical context.
During message passing, the model processes local neighborhoods using attention-based aggregation.
The update rules for node $i$ at $l$-th layer are defined as:
\begin{equation}
\begin{aligned}
\boldsymbol{h}_i^{l+1} &= \boldsymbol{h}_i^{l} + \sum_{j \in N_i} f_{\boldsymbol{h}} (\Vert \boldsymbol{x}_j^l - \boldsymbol{x}_i^l \Vert, \boldsymbol{h}_i^l, \boldsymbol{h}_j^l, \boldsymbol{e}_{ij}), \\
\boldsymbol{x}_i^{l+1} &= \boldsymbol{x}_i^{l} + \sum_{j \in N_i} (\boldsymbol{x}_i^l - \boldsymbol{x}_j^l) f_{\boldsymbol{x}}(\Vert \boldsymbol{x}_j^l - \boldsymbol{x}_i^l \Vert, \boldsymbol{h}_i^{l+1}, \boldsymbol{h}_j^{l+1}, \boldsymbol{e}_{ij}),
\end{aligned}
\end{equation}
where $N_i$ denotes the set of neighboring nodes of node $i$, and $\boldsymbol{e}_{ij}$ denotes the edge features between nodes $i$ and $j$.
The functions $f_{\boldsymbol{h}}$ and $f_{\boldsymbol{x}}$ are implemented as attention-based neural networks that model how feature similarity and geometric distance contribute to updates of invariant and equivariant features, respectively.

The iterative updates of both positional and chemical context features simulate the dynamic interactions between the ligand and the pocket, facilitating the model to capture rich information for the final predictions.

\paragraph{Final Prediction}
Following the complex message passing stage, the learned invariant features $\boldsymbol{h}$ and equivariant coordinates $\boldsymbol{x}$ are utilized to generate the final predictions.

For the ligand generation, the coordinates produced at the final GNN layer are taken directly as the predicted atom positions $\hat{\boldsymbol{x}}_0^{\mathcal{M}}$, while the corresponding hidden features $\boldsymbol{h}^{L,\mathcal{M}}$ are passed through an additional MLP head to output the predicted atom types $\hat{\boldsymbol{v}}_0^{\mathcal{M}}$.

For the protein pocket refinement, we employ a Self-Attention Graph Pooling (SAGPooling) layer~\cite{lee2019self} to aggregate atom-level representations into residue-level features.
Specifically, the concatenated features $[\boldsymbol{x}^{L,\mathcal{P}}, \boldsymbol{h}^{L,\mathcal{P}}]$ are fed into the SAGPooling module, which adaptively selects structurally informative nodes based on learned attention scores. 
The resulting residue-level representations are then passed through separate MLP heads to predict the translation vectors $\hat{\boldsymbol{tr}}_0$, rotation quaternions $\hat{\boldsymbol{q}}_0$, and chi angle updates $\hat{\boldsymbol{\mathcal{X}}}_0$.

To summarize, at each time step $t$, the overall mapping learned by \sysname can be expressed as:
\begin{equation}
(\hat{\boldsymbol{x}}_0^{\mathcal{M}}, \hat{\boldsymbol{v}}_0^{\mathcal{M}}), (\hat{\boldsymbol{tr}}_0, \hat{\boldsymbol{q}}_0, \hat{\boldsymbol{\mathcal{X}}}_0) = \boldsymbol{s}_{\theta}(\boldsymbol{x}_t^{\mathcal{M}}, \boldsymbol{v}_t^{\mathcal{M}}; \boldsymbol{x}_t^{\mathcal{P}}, \boldsymbol{v}_t^{\mathcal{P}}).
\end{equation}

\subsection{Training Objective}
\label{sec:train-obj}
To train \sysname, we approximate the diffusion model to the score function described in \eqn\ref{eqn:pfode}.

The total training objective is composed of multiple loss components, each designed to improve the prediction accuracy of specific outputs for the ligand and the protein pocket.

\paragraph{Ligand Loss}
For the ligand, we incorporate the same objective function as prior work~\cite{guan3d}.
The ligand atom position loss is defined as:
\begin{equation}
    \mathcal{L}_{x}^{\mathcal{M}} = \Vert \hat{\boldsymbol{x}}_0^{\mathcal{M}} - \boldsymbol{x}_0^{\mathcal{M}} \Vert^2.
\end{equation}
To supervise ligand atom types, we compute the KL-divergence of categorical distributions as follows:
\begin{equation}
    \mathcal{L}_{v}^{\mathcal{M}} = \sum_{k}\boldsymbol{c}(\boldsymbol{v}_t^{\mathcal{M}}, \boldsymbol{v}_0^{\mathcal{M}})_k \log\frac{\boldsymbol{c}(\boldsymbol{v}_t^{\mathcal{M}}, \boldsymbol{v}_0^{\mathcal{M}})_k}{\boldsymbol{c}(\boldsymbol{v}_t^{\mathcal{M}}, \hat{\boldsymbol{v}}_0^{\mathcal{M}})_k}.
\end{equation}

\paragraph{Pocket Loss}
For the protein pocket, at each time step $t$, the model is trained to reverse the noisy transformations and recover the original holo state.

The translation loss is defined using mean squared error, where the target is the inverse translation vector obtained via quaternion rotation:
\begin{gather}
    \mathcal{L}_{tr}^{\mathcal{P}} = \Vert \hat{\boldsymbol{tr}}_0 - (-\boldsymbol{q}_0^* \boldsymbol{tr}_0 \boldsymbol{q}_0) \Vert^2.
\end{gather}
For rotation prediction, we apply an L1 loss, along with a unit norm regularization:
\begin{equation}
    \mathcal{L}_{q}^{\mathcal{P}} = |\hat{\boldsymbol{q}}_0 - \boldsymbol{q}_0^*| + (1-\hat{\boldsymbol{q}}_0 \hat{\boldsymbol{q}}_0^*).
\end{equation}
For chi angle updates, we use a cosine-based loss to capture angular differences:
\begin{equation}
    \mathcal{L}_{\mathcal{X}}^{\mathcal{P}} = 1 - \cos(\hat{\boldsymbol{\mathcal{X}}}_0 - (-\boldsymbol{\mathcal{X}}_0)).
\end{equation}

The overall loss function is formulated as a weighted sum of the above five individual loss terms: $\mathcal{L}=\lambda_{1}\mathcal{L}_{x}^{\mathcal{M}} + \lambda_2 \mathcal{L}_{v}^{\mathcal{M}} + \lambda_3 \mathcal{L}_{tr}^{\mathcal{P}} + \lambda_4 \mathcal{L}_{q}^{\mathcal{P}} + \lambda_5 \mathcal{L}_{\mathcal{X}}^{\mathcal{P}}$.

The training and sampling details are summarized in \apx B, where we also provide pseudocode for each procedure.

\section{Experiment}
\label{sec:exp}

\subsection{Experimental Settings}

\paragraph{Dataset Curation}
To evaluate the effectiveness of \sysname, we build a new benchmark dataset of experimental structural pairs of protein–ligand complexes and their corresponding apo conformations.
The PLINDER dataset~\cite{durairaj2024plinder}, which contains more than $400K$ protein–ligand complexes with comprehensive structural annotations, serves as the initial source of structures.
To ensure pharmacologically relevant and high-quality data, we retain only triplets comprising experimentally resolved apo and holo protein structures (with $100\%$ sequence identity and resolution $\leq 2.5$Å) paired with their drug-like small molecule ligands, excluding ions, cofactors, artifacts, molecular fragments, and other undesired ligands.
Additional stringent filtering yields a final benchmark of $24,601$ apo-holo-ligand triplets.
For each complex, the holo pocket is defined as all residues within $10$Å of the ligand atoms, and the corresponding apo pocket is obtained by extracting the same residues from the unbound structure.
The dataset is chronologically split into $23,052$ training, $1,071$ validation, and $478$ test samples to reflect real-world settings.
Full details on the data processing pipeline and statistics are provided in \apx A.

\begin{table*}[t]
\centering
\begin{tabular}{@{}lccccccccccc@{}}
\toprule
\multirow{2}{*}{Method} &
  \multicolumn{2}{c}{Vina min   ($\downarrow$)} &
  \multicolumn{2}{c}{QED ($\uparrow$)} &
  \multicolumn{2}{c}{SA ($\uparrow$)} &
  \multicolumn{2}{c}{Logp} &
  \multicolumn{2}{c}{Linpiski ($\uparrow$)} &
  \multirow{2}{*}{High Affinity ($\uparrow$)} \\ \cmidrule(lr){2-11}
 &
  Avg. &
  Med. &
  Avg. &
  Med. &
  Avg. &
  Med. &
  Avg. &
  Med. &
  Avg. &
  Med. &
  \\ \midrule
Reference &
  -7.41 &
  -7.38 &
  0.61 &
  0.64 &
  0.71 &
  0.71 &
  2.08 &
  2.56 &
  4.67 &
  5.00 &
  \textbackslash{} 
   \\ \midrule
IPDiff &
  \underline{-6.40} &
  \underline{-6.56} &
  0.51 &
  0.52 &
  0.60 &
  0.60 &
  1.90 &
  2.27 &
  4.65 &
  5.00 &
  29.6\%
   \\
TargetDiff &
  -5.19 &
  -5.20 &
  0.37 &
  0.38 &
  0.57 &
  0.57 &
  1.69 &
  1.52 &
  4.68 &
  5.00 &
  33.8\%
   \\
Pocket2Mol &
  -3.30 &
  -3.29 &
  0.46 &
  0.46 &
  \textbf{0.85} &
  \textbf{0.86} &
  0.83 &
  1.02 &
  \textbf{4.98} &
  5.00 &
  \textbackslash{}
   \\
DecompDiff &
  -6.37 &
  -6.40 &
  \underline{0.56} &
  \underline{0.58} &
  \underline{0.66} &
  \underline{0.66} &
  1.53 &
  1.76 &
  4.73 &
  5.00 &
  \underline{34.3\%}
   \\ \midrule
Apo2Mol - Gen &
  \textbf{-6.79} &
  \textbf{-7.09} &
  \textbf{0.59} &
  \textbf{0.63} &
  0.61 &
  0.61 &
  3.25 &
  3.04 &
  \underline{4.78} &
  5.00 &
  \textbf{42.7\%}
   \\ \bottomrule
\end{tabular}
\caption{Benchmark on ligand generation from \textbf{apo structures}. Baseline models are evaluated on apo structures, while \sysname generates both refined pockets and novel ligands from the apo inputs, with ligand accessed against the generated pockets. The best and second results are highlighted in \textbf{bold} and \underline{underlined}. This convention is used throughout subsequent tables.}
\label{tab:benchmark-apo}
\end{table*}

\begin{table*}[t]
\centering
\begin{tabular}{@{}lccccccccccc@{}}
\toprule
\multirow{2}{*}{Method} &
  \multicolumn{2}{c}{Vina min   ($\downarrow$)} &
  \multicolumn{2}{c}{QED ($\uparrow$)} &
  \multicolumn{2}{c}{SA ($\uparrow$)} &
  \multicolumn{2}{c}{Logp} &
  \multicolumn{2}{c}{Linpiski ($\uparrow$)} &
  \multirow{2}{*}{High Affinity ($\uparrow$)} \\ \cmidrule(lr){2-11}
 &
  Avg. &
  Med. &
  Avg. &
  Med. &
  Avg. &
  Med. &
  Avg. &
  Med. &
  Avg. &
  Med. &
 \\ \midrule
Reference &
  -7.41 &
  -7.38 &
  0.61 &
  0.64 &
  0.71 &
  0.71 &
  2.08 &
  2.56 &
  4.67 &
  5.00 &
  \textbackslash{}
   \\ \midrule
IPDiff &
  \underline{-7.09} &
  \underline{-7.08} &
  0.52 &
  0.54 &
  0.61 &
  0.61 &
  2.17 &
  2.42 &
  4.67 &
  5.00 &
  \underline{44.9\%}
   \\
TargetDiff &
  -5.50 &
  -5.56 &
  0.39 &
  0.39 &
  0.57 &
  0.57 &
  1.79 &
  1.70 &
  4.71 &
  5.00 &
  35.1\%
   \\
Pocket2Mol &
  -3.32 &
  -3.35 &
  0.46 &
  0.46 &
  \textbf{0.87} &
  \textbf{0.88} &
  0.92 &
  1.12 &
  \textbf{4.99} &
  5.00 &
  \textbackslash{}
   \\
DecompDiff &
  -6.39 &
  -6.41 &
  \underline{0.56} &
  \underline{0.58} &
  \underline{0.65} &
  \underline{0.65} &
  1.53 &
  1.76 &
  4.74 &
  5.00 &
  31.9\%
   \\ \midrule
Apo2Mol - holo &
  \textbf{-7.86} &
  \textbf{-8.03} &
  \textbf{0.61} &
  \textbf{0.65} &
  0.59 &
  0.62 &
  3.49 &
  3.35 &
  \underline{4.76} &
  5.00 &
  \textbf{52.9\%}
   \\ \bottomrule
\end{tabular}
\caption{Benchmark on ligand generation from \textbf{holo structures}. Baseline models are evaluated on native holo structures, while \sysname generates novel ligands directly from the apo inputs, with ligand assessed against the native holo pockets.}
\label{tab:benchmark-holo}
\end{table*}

\paragraph{Baseline Methods}
We compare \sysname against four representative baseline methods in SBDD:
1) \textbf{Pocket2Mol}~\cite{Pocket2Mol}: a generative model that constructs 3D molecules by sequentially placing atoms around a given protein pocket.
2) \textbf{TargetDiff}~\cite{guan3d}: a diffusion-based method that generates ligand atom coordinates and types, with bond information determined via a post-processing step.
3) \textbf{DecompDiff}~\cite{guan2024decompdiff}: a diffusion model that decomposes ligands into scaffold and arm substructures and diffuses them separately using bond-level guidance and structural priors.
4) \textbf{IPDiff}~\cite{huang2024protein}: it integrates protein–ligand interaction priors into both the forward and reverse diffusion processes to improve binding-aware generation.

While FlexSBDD~\cite{zhang2024flexsbdd} and DynamicFlow~\cite{zhou2025integrating} provide alternative frameworks for incorporating pocket dynamics into molecule generation, we exclude them from benchmarking since their implementations are not publicly available, preventing reproducibility.

\paragraph{Evaluation Metrics}
To evaluate the quality of the generated ligands, we consider six widely used metrics. 1) \textbf{Binding Affinity}, measured by the Vina min score computed using AutoDock Vina~\cite{eberhardt2021autodock} under standard settings following prior work~\cite{zhou2025integrating}. 2) \textbf{QED}~\cite{bickerton2012quantifying}, a quantitative estimate of drug-likeness. 3) \textbf{SA Score}~\cite{ertl2009estimation}, reflecting synthetic accessibility. 4) \textbf{logP}~\cite{ghose1999knowledge}, the octanol–water partition coefficient; values between $-0.4$ and $5.6$ indicate favorable pharmacokinetic properties. 5) \textbf{Lipinski}~\cite{lipinski2012experimental}, denoting the number of molecular properties that satisfy Lipinski’s Rule of Five (\apx A). 6) \textbf{High Affinity} measures the percentage of generated ligands with binding affinity higher than the reference molecules.
All evaluation metrics, except High Affinity, are reported using both their mean and median values.

To assess the plausibility of the generated pockets, we compute the RMSD to their apo counterparts and calculate the Jensen–Shannon Divergence (JSD) between RMSD distributions of generated and experimental holo pockets, quantifying how well the model captures apo-to-holo transformations.

\begin{figure}[!t]
    \centering
    \includegraphics[width=0.4\textwidth]{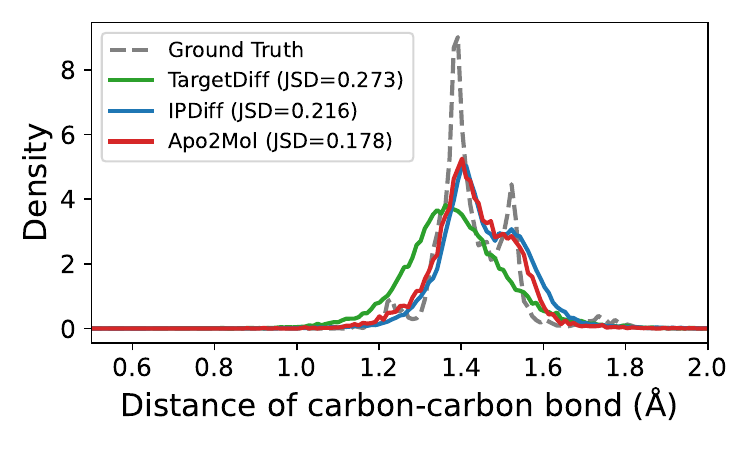}
    \caption{
        Distribution comparison for distances of carbon-carbon pairs for ground truth molecules in the test set (gray) and model-generated molecules (color). JSD between two distributions is reported.
    }
    \label{fig:bond-dist}
\end{figure}

\begin{figure}[!t]
    \centering
    \includegraphics[width=0.4\textwidth]{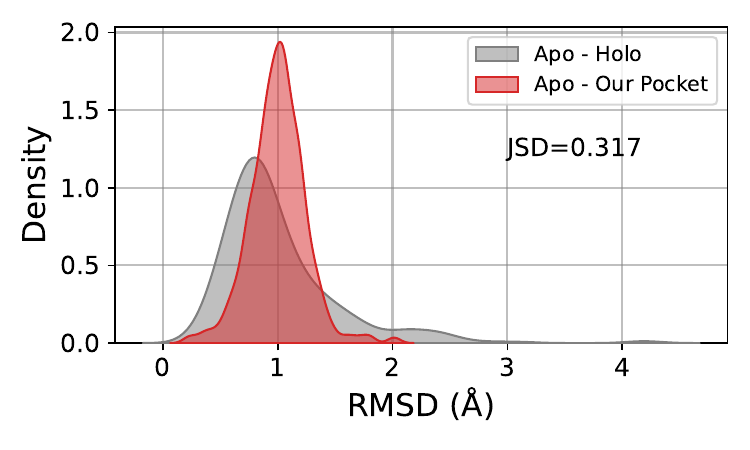}
    \caption{
        RMSD distribution between apo and holo, and between apo and generated pockets. JSD denotes the Jensen–Shannon divergence between the two distributions.
    }
    \label{fig:rmsd-distribution}
\end{figure}

\begin{figure*}[!t]
    \centering
    \includegraphics[width=0.85\textwidth]{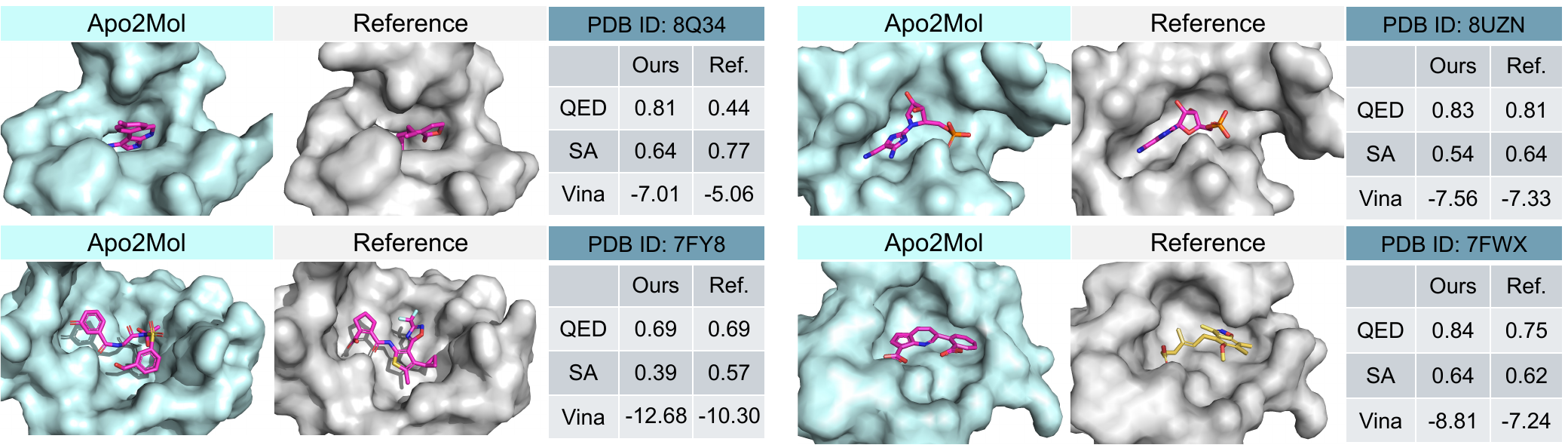}
    \caption{
        Visualization of ligands and pockets generated by \sysname, compared with the corresponding reference ligands and holo structures. QED, SA, and Vina min scores are reported.
    }
    \label{fig:vis-example}
\end{figure*}

\subsection{Main Results}

\paragraph{Binding Affinity and Molecule Properties}
To evaluate the effectiveness of \sysname in ligand generation with dynamic pocket refinement, we conduct two sets of comparative experiments.
In the first setting, all baseline models are trained on holo pockets following their original protocols.
During evaluation, only apo conformations are provided to both the baselines and \sysname, simulating a realistic scenario in which the target protein lacks an experimentally determined ligand-bound state.
Since \sysname jointly generates both a refined pocket and a ligand, its ligand evaluation is performed on the generated pocket conformation.
As shown in \tab\ref{tab:benchmark-apo}, \sysname achieves the best performance across key metrics: the average Vina min score improves from $-6.40$ (best baseline, IPDiff) to $-6.79$, and the median score increases from $-6.56$ to $-7.09$, indicating a strong shift toward higher-affinity binding.
\sysname also outperforms baselines on molecular property metrics like QED, suggesting that refining pockets from the apo to holo state enables the generation of ligands with more drug-like profiles.

In the second setting, all baselines are trained and evaluated directly on holo pockets, following the previous \textit{de novo} design setup disregarding protein conformational dynamics.
By contrast, \sysname continues to generate the ligands from the apo inputs, with ligand quality assessed against the corresponding native holo pocket.
As shown in \tab~\ref{tab:benchmark-holo}, although the baselines perform better under this idealized setting than the previous one, \sysname remains competitive, achieving a substantial gain in binding affinity with an average Vina min score of $-7.86$ versus $-7.09$ for the second best model IPDiff, and a median of $-8.03$ versus $-7.08$.
In addition, \sysname also attains a higher High Affinity rate of $52.9\%$, compared with $44.9\%$ for IPDiff.

\begin{table}[t]
\centering
\begin{tabular}{@{}lcccc@{}}
\toprule
\multirow{2}{*}{Method} & \multicolumn{2}{c}{Vina min   ($\downarrow$)} & \multicolumn{2}{c}{QED ($\uparrow$)} \\ \cmidrule(l){2-5} 
 & Avg. & Med. & Avg. & Med. \\ \midrule
\sysname & \textbf{-6.79} & \textbf{-7.09} & \textbf{0.587} & \textbf{0.629} \\
w/o complex graph & -6.18 & -6.22 & \underline{0.524} & 0.529 \\
w/o quaternion & \underline{-6.51} & \underline{-6.44} & 0.523 & \underline{0.534} \\ \bottomrule
\end{tabular}
\caption{Ablation study on key network components.}
\label{tab:ablation}
\end{table}

\paragraph{Molecule Structure}
We assess the structural plausibility of \sysname-generated molecules by comparing carbon–carbon bond distance distributions.
As shown in \fig\ref{fig:bond-dist}, \sysname achieves the lowest JSD ($0.178$) to ground truth, outperforming IPDiff ($0.216$) and TargetDiff ($0.273$), demonstrating its ability to generate more realistic molecular structures.

\paragraph{Pocket Structure}
We compare the RMSD distributions of experimental apo–holo pocket pairs with those of apo–\sysname-generated pockets to evaluate how effectively \sysname reproduces realistic ligand-binding pocket conformations.
As shown in \fig\ref{fig:rmsd-distribution}, the generated pockets generally follow the apo–holo transition distribution (\apx A) but exhibit a mild shift (JSD = $0.317$), suggesting a potential limitation of \sysname in fully modeling the conformational variability of individual apo pockets.

Taken together, these results demonstrate that by jointly modeling ligand generation, pocket dynamics, and ligand–pocket interactions within a unified diffusion framework, \sysname can produce high-quality drug-like ligands and corresponding ligand-pocket conformations even in the absence of holo structures, offering a robust solution to real-world SBDD scenarios where only apo structures are available.

\subsection{Ablation Study}
We conduct ablations to evaluate the contribution of key components in \sysname.
As shown in \tab\ref{tab:ablation}, removing the complex graph—by using only a single edge type within pocket atoms, leads to notable drops in both binding affinity and drug-likeness (Vina min: $-6.18$; QED Avg: $0.524$).
Replacing quaternions with rotation vectors for conformational modeling also degrades performance (Vina min: $-6.51$; QED Avg: $0.523$).
These results highlight the effectiveness of both the hierarchical graph design and quaternion-based transformations in generating high-quality ligand–pocket structures.

\subsection{Visualization}
\fig\ref{fig:vis-example} illustrates four representative examples of ligands and pockets generated by \sysname, shown alongside the corresponding reference ligands and holo structures.
Across diverse targets, \sysname generates ligands with favorable binding affinity and drug-likeness, while also recovering plausible pocket geometries.
These visualizations further highlight the ability of \sysname to produce chemically and structurally realistic complexes directly from apo inputs.

\section{Conclusion and Limitation}
\label{sec:conclusion-limitation}
We present \sysname, a full-atom SE(3)-equivariant diffusion framework that jointly generates small molecule ligands and refined protein pockets directly from apo states.
Through extensive evaluations, \sysname demonstrates strong performance compared to existing baselines and effectively models the protein dynamics during the ligand \textit{de novo} design process.
Despite these advances, a modest distribution gap remains between generated and native holo pockets, suggesting that additional refinement is needed to fully capture fine-grained conformational transitions for specific protein targets.
Future work may benefit from pretraining on diverse protein structures to further close this gap.

\newpage

\section*{Acknowledgement}
This work was supported in part by the University of Florida (UF Startup Fund, UF Health Cancer Institute Pilot Grant \#UFS-2023-08, and UF Research AI Award to Y. L.), National Institutes of Health (R21EB037868 to Y. L.), and the Bodor Professorship Fund (to C. L.). We acknowledge UFIT Research Computing for providing computational resources.

\bibliography{aaai2026}

\newpage
\twocolumn[{
\begin{center}
    \vspace{1em}
    \textbf{\LARGE Appendix}
    \vspace{2em}
\end{center}
}]

\appendix
\setcounter{secnumdepth}{2}

\section{Dataset Construction}
\label{apx:data-cons}

\subsection{Introduction of PLINDER Dataset}
High-quality protein–ligand interaction (PLI) data are fundamental to the development of data-driven approaches to advance structure-based drug design~\cite{durant2024future}.
Recent datasets, such as CrossDocked~\cite{francoeur2020three}, MISATO~\cite{siebenmorgen2024misato}, DecoyDB~\cite{zhang2025decoydb}, SAIR~\cite{lemos2025sair}, and GatorAffinity-DB~\cite{wei2025gatoraffinity} have significantly expanded the available PLI landscape by incorporating computationally generated structures derived from protein-ligand co-folding, docking, and molecular dynamics simulations. 
However, these computational approaches may introduce noise or artificial biases stemming from their suboptimal accuracy and methodological constraints~\cite{scardino2023good,masters2025investigating, khiari2025synthetic}.
Moreover, characterizing the conformational dynamics of protein binding pockets requires datasets that contain both unbound (apo) and ligand-bound (holo) protein structures.
The development of the PLINDER~\cite{durairaj2024plinder} dataset helps eliminate these limitations by offering a large collection of experimentally resolved apo–holo structure pairs.

Constructed from the recent Protein Data Bank (PDB) releases as of 2024-04-09, PLINDER encompasses a total of $449,383$ PLI systems.
In addition to holo structures, it provides  $98,473$ experimentally resolved apo structures and $205,300$ AlphaFold2-predicted structures~\cite{jumper2021highly}, each linked to corresponding holo entries.
PLINDER also includes comprehensive ligand annotations, such as QED score~\cite{bickerton2012quantifying} and CCD code~\cite{westbrook2015chemical}, enabling efficient and flexible task-specific filtering.

\begin{table}[ht]
\centering
\begin{tabular}{@{}ll@{}}
\toprule
\textbf{Annotation} & \textbf{Filtering Rule} \\ \midrule
entry\_resolution & $\leq 2.5$ \\
ligand\_qed & $\geq 0.2$ \\
ligand\_is\_ion & Removed \\
ligand\_is\_fragment & Removed \\
ligand\_is\_oligo & Removed \\
ligand\_is\_cofactor & Kept \\
ligand\_in\_artifact\_list & Removed \\
ligand\_is\_artifact & Removed \\
ligand\_is\_other & Removed \\
ligand\_is\_invalid & Removed \\
ligand\_unique\_ccd\_code & Removed \\
ligand\_is\_proper & Kept \\ 
\bottomrule
\end{tabular}
\caption{Filtering Criteria for Drug-like Ligands}
\label{tab:ligand-filter}
\end{table}

\subsection{Data Filtering}
To ensure high-quality data for training and evaluation, we first select samples from the PLINDER dataset in which holo protein structures are paired with experimentally resolved apo structures, excluding AlphaFold2-predicted models to avoid potential noise introduced by structure prediction inaccuracies.
To guarantee strict structural matching, we retain only apo-holo pairs with $100\%$ sequence identity.
We then filter for drug-like ligands using molecular property annotations provided by PLINDER, with the specific filtering criteria summarized in \tab\ref{tab:ligand-filter}.
Ligands containing metal elements ($4.3\%$ of the total) are further excluded.

Our final curated dataset contains $24,601$ apo-holo-ligand triplets.
Compared with ApoBind~\cite{aggarwal2021apobind}, which contains $\sim10K$ samples with $\ge80\%$ apo-holo sequence identity, our dataset is over twice as large, and ensures residue-level correspondence between apo and holo structures, making it more suitable for the requirements of this study.

For dataset splitting, we adopt the recommended chronological partitioning strategy in PLINDER (also used in DynamicBind~\cite{lu2024dynamicbind}) to better reflect real-world deployment scenarios.
To ensure strict structural non-overlap, we further verify that every test pocket has TM-scores~\cite{zhang2004scoring} $<0.35$ to all training pockets.
All baselines are trained and evaluated using this same split.
The resulting benchmarking contains $23,052$ training examples, $1,071$ validation examples, and $478$ test examples.

\subsection{Pocket Generation}
Since \sysname generates novel small molecule ligands for specific protein pockets, we follow established practices~\cite{guan3d} for pocket extraction from holo protein structures.
The holo pocket is defined by all amino acids within a $10$Å radius of any of the bound ligand atom.
The corresponding apo pocket is then obtained by extracting the same residues from the apo structure.

\begin{figure}[t]
    \centering
    \includegraphics[width=0.47\textwidth]{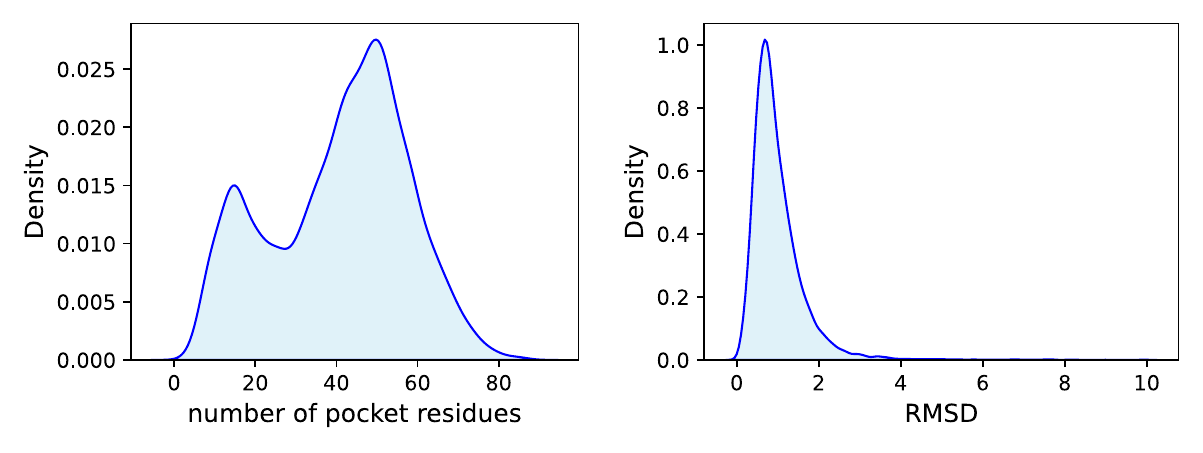}
    \caption{Statistical results of the protein pockets in \sysname dataset. \textbf{RMSD} represents the RMSD between the Holo pocket and the Apo pocket.}
    \label{fig:pocket-stat}
\end{figure}

\begin{figure}[t]
    \centering
    \includegraphics[width=0.3\textwidth]{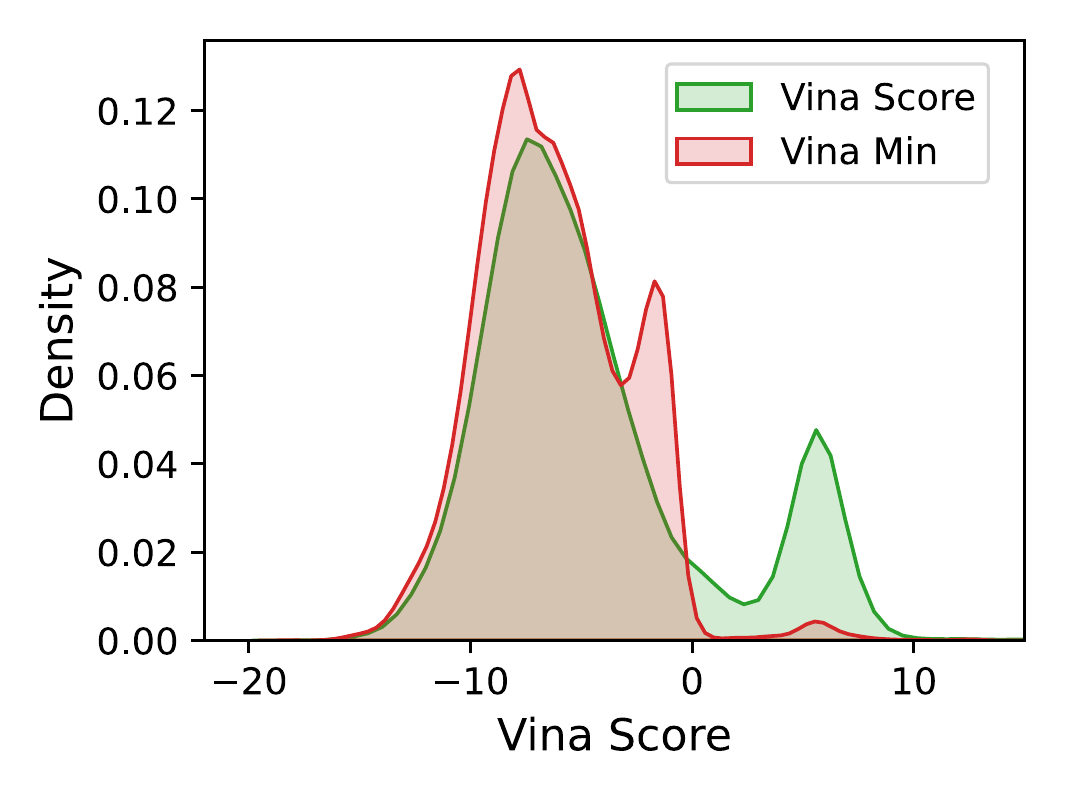}
    \caption{Statistical results of the Vina score and the Vina min in \sysname dataset. The results are calculated with the holo protein pockets.}
    \label{fig:vina-stat}
\end{figure}

\begin{figure}[t]
    \centering
    \includegraphics[width=0.47\textwidth]{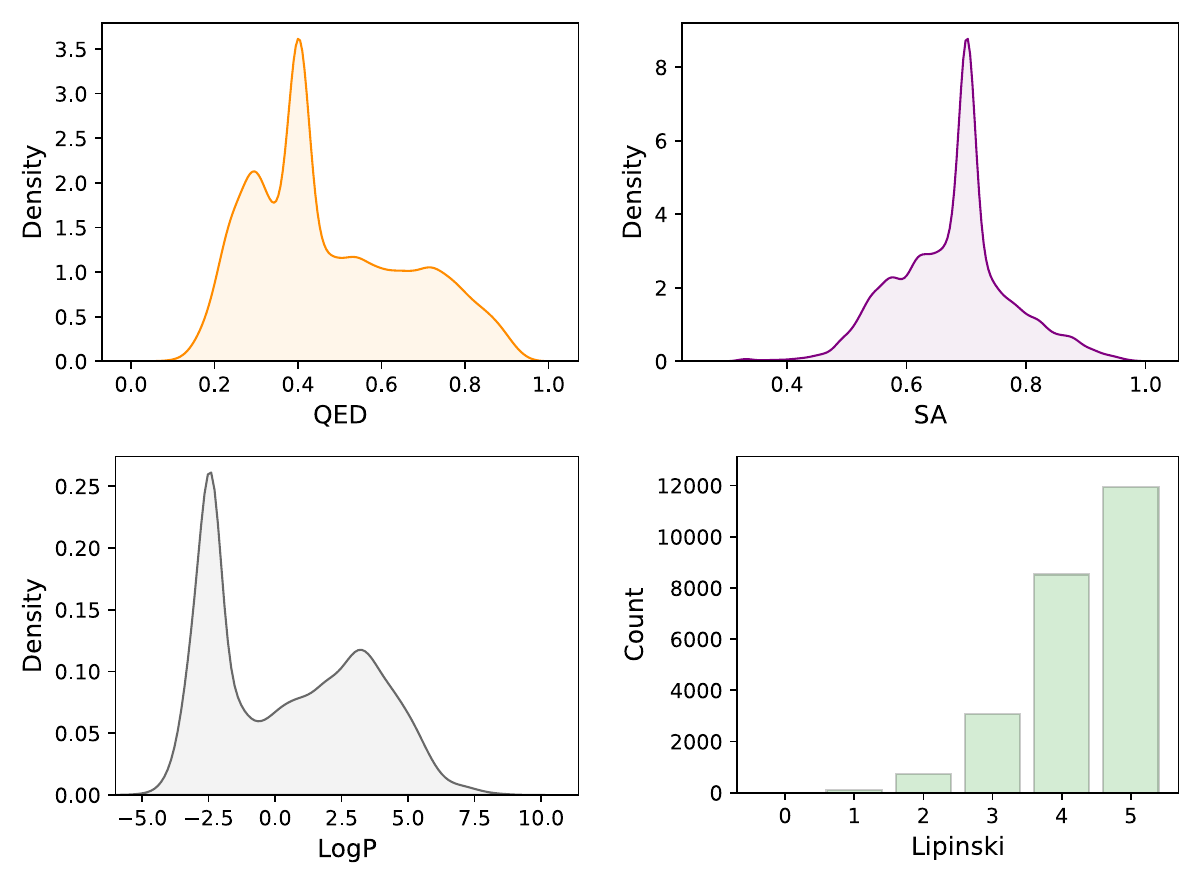}
    \caption{Statistical results of the reference ligand properties, including QED, SA, logP, and Lipinski.}
    \label{fig:ligand-stat}
\end{figure}

\subsection{Statistical Analysis}

\fig~\ref{fig:pocket-stat} summarizes the key statistics of protein pockets in the \sysname dataset.
The left panel shows the distribution of residue counts per pocket, while the right panel plots the RMSD distribution of apo–holo pocket pairs.
The binding affinities of reference ligands to their native holo pockets are measured using Vina score and Vina min.
The Vina score is calculated directly from the given ligand and protein pocket structures, whereas Vina min performs an additional local structure energy minimization on the ligand conformations before estimation.
The results are shown in \fig\ref{fig:vina-stat}.

\fig\ref{fig:ligand-stat} illustrates the distributions of ligand properties, including QED, SA, logP, and Lipinski.
Specifically, the Lipinski score represents the number of satisfied criteria among the following five rules:
\begin{enumerate}
    \item molecular weight $< 500$ Da;
    \item number of hydrogen bond donors $\le 5$;
    \item number of hydrogen bond acceptors $\le 10$;
    \item logP value within the range $[-2, 5]$;
    \item number of rotatable bonds $\le 10$.
\end{enumerate}

\section{Implementation}
\label{apx:implementation}

\subsection{Computational Resource}
\sysname is trained using the Adam optimizer implemented in PyTorch on four NVIDIA A100-80G SXM4 GPUs.
The initial learning rate is set to $5\times10^{-4}$ and decayed using a plateau scheduler with a decay factor of 0.75, a patience of 10 epochs, and a minimum learning rate of $1\times10^{-6}$.
We use a batch size of 2 per GPU, resulting in a total batch size of 8.
Training typically converges within 150 epochs.

The training hyperparameters for \sysname are summarized in \tab\ref{tab:hyperparameter}.
For all baseline methods, we follow their original training configurations to ensure consistency.

\subsection{Training \& Sampling}
We provide the training and sampling procedure in pseudo-code in \alg\ref{alg:training} and \alg\ref{alg:sampling}, respectively.

\begin{algorithm}
\caption{Training Procedure of \sysname}
\label{alg:training}
\begin{algorithmic}[1]
\Require Dataset $\mathcal{D} = \{(\mathcal{P}_{\text{apo}}, \mathcal{P}_{\text{holo}}, \mathcal{M})\}$; diffusion steps $T$; model $\boldsymbol{s}_{\theta}$
\For{each training iteration}
    \State Sample $(\mathcal{P}_{\text{apo}}, \mathcal{P}_{\text{holo}}, \mathcal{M}) \sim \mathcal{D}$
    \State Sample timestep $t \sim \text{Uniform}(\{1, \dots, T\})$
    \State Compute residue-level transformations $(\boldsymbol{t}_r, \boldsymbol{q}, \boldsymbol{\mathcal{X}})$ from $\mathcal{P}_{\text{holo}} \rightarrow \mathcal{P}_{\text{apo}}$
    \State Interpolate to obtain pseudo pocket $\mathcal{P}_t$ using $(\boldsymbol{t}_r^{(t)}, \boldsymbol{q}^{(t)}, \boldsymbol{\mathcal{X}}^{(t)})$ \Comment{See Eq.~\ref{eqn:tr-x-t},~\ref{eqn:q-t}}
    \State Apply forward diffusion to ligand positions and atom types to obtain $\mathcal{M}_t$ \Comment{See Eq.~\ref{eqn:q-t}}
    \State Encode $\mathcal{P}_t$ and $\mathcal{M}_t$ using pocket and ligand encoders
    \State Build a cross-modal graph with hierarchical message passing
    \State Predict denoised outputs:
    \[
    (\hat{\boldsymbol{x}}_0^{\mathcal{M}}, \hat{\boldsymbol{v}}_0^{\mathcal{M}}), (\hat{\boldsymbol{tr}}_0, \hat{\boldsymbol{q}}_0, \hat{\boldsymbol{\mathcal{X}}}_0) = \boldsymbol{s}_{\theta}(\boldsymbol{x}_t^{\mathcal{M}}, \boldsymbol{v}_t^{\mathcal{M}}; \boldsymbol{x}_t^{\mathcal{P}}, \boldsymbol{v}_t^{\mathcal{P}})
    \]
    \State Compute total loss $\mathcal{L}$ \Comment{See Sec.~\ref{sec:train-obj}}
    \State Update model parameters $\theta$ via gradient descent
\EndFor
\end{algorithmic}
\end{algorithm}

\begin{algorithm}
\caption{Generation Procedure of \sysname}
\label{alg:sampling}
\begin{algorithmic}[1]
\Require Apo pocket structure $\mathcal{P}_{\text{apo}}$; number of diffusion steps $T$; trained model $\boldsymbol{s}_{\theta}$
\State Initialize ligand atom positions $\boldsymbol{x}_T^{\mathcal{M}} \sim \mathcal{N}(0, I)$ and types $\boldsymbol{v}_T^{\mathcal{M}} \sim \text{Uniform}([1, K])$
\State Set initial pocket conformation $\mathcal{P}_T \gets \mathcal{P}_{\text{apo}}$
\For{$t = T, T-1, \dots, 1$}
    \State Encode $\mathcal{P}_t$ and $\mathcal{M}_t$
    \State Construct a complex graph and apply hierarchical message passing
    \State Predict denoised outputs:
    \[
    (\hat{\boldsymbol{x}}_0^{\mathcal{M}}, \hat{\boldsymbol{v}}_0^{\mathcal{M}}), (\hat{\boldsymbol{tr}}_0, \hat{\boldsymbol{q}}_0, \hat{\boldsymbol{\mathcal{X}}}_0) = \boldsymbol{s}_{\theta}(\boldsymbol{x}_t^{\mathcal{M}}, \boldsymbol{v}_t^{\mathcal{M}}; \boldsymbol{x}_t^{\mathcal{P}}, \boldsymbol{v}_t^{\mathcal{P}})
    \]
    \State Sample ligand and pocket at time step $t-1$ using posterior distribution
\EndFor
\State \Return $(\hat{\boldsymbol{x}}_0^{\mathcal{M}}, \hat{\boldsymbol{v}}_0^{\mathcal{M}})$, predicted ligand; $(\hat{\boldsymbol{tr}}_0, \hat{\boldsymbol{q}}_0, \hat{\boldsymbol{\mathcal{X}}}_0)$, predicted Holo pocket transformation
\end{algorithmic}
\end{algorithm}

\begin{table*}[t]
\centering
\begin{tabular}{@{}lcl@{}}
\toprule
Hyperparameter & Value & Description \\ \midrule
hidden\_dim & 128 & hidden dimention in \sysname \\
n\_layers & 9 & number of GNN layers in \sysname \\
k\_neighbors & 32 & number of neighbors for each node in the KNN graph \\
beta\_start & 1.e-7 & beta start value for the diffusion model \\
beta\_end & 2.e-3 & beta end value for the diffusion model \\
sampling steps & 1000 & number of sampling steps in the diffusion model \\ \bottomrule
\end{tabular}
\caption{Hyperparameters for \sysname}
\label{tab:hyperparameter}
\end{table*}

\section{Additional Experiment}
\label{apx:add-exp}

\subsection{Ligand Analysis}
\label{apx:ligand-analysis}

In addition to the six primary evaluation metrics reported in the main text, we further evaluate the validity and novelty of the model-generated ligands, where novelty is defined as a Tanimoto similarity $\leq 0.4$ (computed using Morgan fingerprints) relative to the reference ligands.
We compare \sysname with the strongest baseline, IPDiff, and for each target, both methods generate five ligands under the first experimental setting, where ligand generation is performed directly from apo states.
As shown in \tab\ref{tab:ligand-generalization}, \sysname achieves better performances in both evaluation metrics.

\begin{table}[h]
\centering
\begin{tabular}{lcc}
\toprule
\textbf{Method} & \textbf{Validity $\uparrow$} & \textbf{Novelty $\uparrow$} \\
\midrule
\sysname & \textbf{88.9\%} & \textbf{95.3\%} \\
IPDiff  & 87.6\% & 91.1\% \\
\bottomrule
\end{tabular}
\caption{Generalization comparison of ligand quality.}
\label{tab:ligand-generalization}
\end{table}

\begin{figure}[h]
    \centering
    \includegraphics[width=0.46\textwidth]{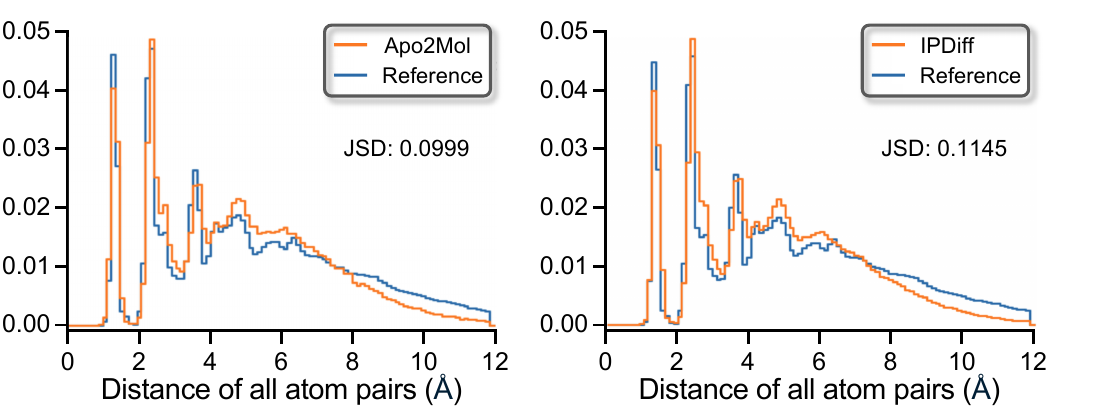}
    \caption{Distance distribution of all atom pairs for reference molecules (blue) and model-generated molecules (orange). Jensen–Shannon divergence (JSD) is reported for each method.}
    \label{fig:all-atom-dis}
\end{figure}

To further characterize the model-generated molecular conformation, we analyze the all-atom pairwise distance distributions of ligands generated by \sysname and IPDiff, and compare them with reference molecules in the test set.
As shown in \fig\ref{fig:all-atom-dis}, \sysname achieves a lower Jensen–Shannon divergence from the reference distribution than IPDiff, indicating better alignment with the underlying structural statistics.

\subsection{Pocket Analysis}
\label{apx:pocket-analysis}

We further examine pocket-level geometric consistency.
The volume distribution of \sysname-generated pockets closely matches the holo references, achieving a JSD of $0.010$, compared with $0.022$ for the input Apo pockets.
Nonetheless, we note that volume agreement alone does not guarantee full physical plausibility of the generated pocket conformations.
Developing more principled metrics for assessing pocket-shape realism will be an important direction for future work.

\end{document}